\documentclass[usenatbib]{mnras}
\usepackage[utf8]{inputenc}
\usepackage{hyperref}
\usepackage{graphicx}
\usepackage{mathtools}
\usepackage{amsmath}
\usepackage{widetext}
\usepackage{xcolor}
\usepackage{orcidlink}
\usepackage{newtxtext,newtxmath}
\usepackage[T1]{fontenc}
\usepackage{ae,aecompl}

\definecolor{dg}{rgb}{0.0, 0.6, 0.1}

\newcommand{\Myr}{Myr}
\newcommand{\He}{He}
\newcommand{\Fe}{Fe}
\newcommand{\Ox}{O}
\newcommand{\halo}{scattering halo}
\newcommand{\haloes}{scattering haloes}
\newcommand{\skymapdir}{figures}

\date{}

\title{UHECR Echoes from the Council of Giants}

\author[Taylor, Matthews \& Bell]{
A.~M.~Taylor \orcidlink{0000-0001-9473-4758},$^{1}$\thanks{E-mail: andrew.taylor@desy.de}
J.~H.~Matthews \orcidlink{0000-0002-3493-7737}$^{2,3}$ and
A.~R.~Bell \orcidlink{0000-0002-8843-5003}$^{4,5}$
\\
$^{1}$Deutsches Elektronen-Synchrotron, Platanenallee 6, Zeuthen, Germany \\ 
$^{2}$Department of Physics, Astrophysics, University of Oxford, Denys Wilkinson Building, Keble Road, Oxford, OX1 3RH, UK\\
$^{3}$Institute of Astronomy, University of Cambridge, Madingley Road, Cambridge, CB3 0HA, UK\\
$^{4}$University of Oxford, Clarendon Laboratory, Parks Road, Oxford, OX1 3PU, UK\\
$^{5}$Central Laser Facility, STFC Rutherford Appleton Laboratory, Harwell, Oxford,  OX11 0QX, UK
}

\date{\today}

\pubyear{2023}

\begin{document}
\label{firstpage}
\pagerange{\pageref{firstpage}--\pageref{lastpage}}
\maketitle

\begin{abstract}
Recent anisotropy studies of UHECR data at energies $\gtrsim 40 {\rm ~EeV}$, have disclosed a correlation of their angular distribution with the extragalactic local structure, specifically with either local starburst galaxies or AGN. Using Monte Carlo simulations taking into account photo-disintegration processes, we further explore a framework in which these UHECRs were accelerated by Centaurus~A in a recent powerful outburst before being scattered by magnetic fields associated with local, Council of Giant, extragalactic structure.  
We find that the observed intermediate scale anisotropies can be accounted for by the Council of Giant structure imposing a response function on the initial outburst of UHECRs from a single source located at Centaurus~A's position. The presence of these local structures creates `echoes' of UHECRs after the initial impulse, and focusing effects. The strongest echo wave has a lag of $\sim 20$~\Myr, comparable to the age of synchrotron-emitting electrons in the giant Centaurus~A lobes.  Through consideration of the composition of both the direct and echo wave components, we find that the distribution of the light ($1<\ln A<1.5$) component across the sky offers exciting prospects for testing the echo model using future facilities such as Auger prime. Our results demonstrate the potential that UHECR nuclei offer, as “composition clocks”, for probing  propagation scenarios from local sources. 
\end{abstract}
\bigskip

\section{Introduction}
\label{intro}

The question as to the origin of the highest energy cosmic rays, with energies in excess of $10^{20}$~eV, which have been detected at Earth over the past 60~years \citep{1963PhRvL..10..146L}, continues to drive observational and theoretical studies in high energy astrophysics. Despite the time passed since their first detection, the answer to this question remains unresolved. 

On theoretical grounds, the Hillas criterion \citep{hillas_origin_1984} and Hillas-Lovelace condition indicate that the most promising candidates are objects possessing fast outflows with high kinetic energy luminosities  \citep{,lovelace_dynamo_1976,2004Prama..62..483W,1995ApJ...454...60N,blandford_acceleration_2000} such as active galactic nuclei (AGN) and gamma-ray bursts (GRB). Additionally, the limited propagation distance of ultra high energy cosmic ray (UHECR) nuclei through extragalactic radiation fields further constrains the number of potential candidate objects, to only those in the relatively local extragalactic vicinity \citep{Taylor:2011ta,lang_revisiting_2020}. For AGN, only a few local candidate sources exist, such as Centaurus~A (Cen~A) \citep{OSullivan:2009rvg,2009A&A...506L..41R}. 

Recently, new insights into this UHECR origins problem have been provided by the Pierre Auger Observatory (PAO), which has reported a correlation of the UHECR hotspots seen in their skymaps with local structure in the southern hemisphere sky, specifically with either nearby star-forming galaxies (which they referred to as starburst galaxies) or AGN \citep{2018ApJ...853L..29A,PierreAuger:2022axr}. Likewise, similar correlations of UHECR hotspots, for energies above 40~EeV, with local structure in the northern hemisphere sky have been reported by the Telescope Array (TA) collaboration \citep{TelescopeArray:2014tsd}. With the significance of the PAO reported starburst galaxy correlation being already larger than $4~\sigma$ (post-trial), the origin of such a correlation appears worthy of deeper consideration.

\defcitealias{bell_echoes_2022}{BM22}
The existence of this correlation, assuming that it is a real correlation and not simply a statistical fluctuation, raises the question as to whether such a correlation can be compatible with a scenario in which a local AGN, namely Cen~A, is the source of the UHECR driving the anisotropy signal detected by the PAO. We here explore the possibility that a correlation of UHECR with local structure is brought about by the deflection of UHECR, initially released by Cen~A, on nearby galaxy systems, a question first raised by \citet[][hereafter \citetalias{bell_echoes_2022}]{bell_echoes_2022}.

In section~\ref{local_structure} we consider the Milky Way's local extragalactic neighbourhood. In section~\ref{sim_setup} we describe the setup considered to study the propagation of UHECR from Cen~A to Earth, considering their scattering the magnetic field associated with local galaxies, and their energy- and species- dependent photo-disintegration in extragalactic radiation fields. In section~\ref{results} the key findings from our simulations are discussed.
In section~\ref{discussion} we discuss these results, outlining the limitations of our approach and indicating further aspects to be explored. In section~\ref{conclusion} we draw our conclusions.

\section{The Local Extragalactic Environment}
\label{local_structure}

Following the growth of structure formation via gravitational collapse over cosmological timescales, the Universe at the present epoch on small scales ($\lesssim100$~Mpc), is inhomogeneous. In the current study we zoom in on the inhomogeneous patch of the Universe in which the Milky Way (MW) resides. Specifically, we focus here on very local distances $\lesssim10$~Mpc around the MW, in a region with distinct kinematics known as the Local Sheet \citep{tully_our_2008}. The most massive galaxies in this region form a ring approximately surrounding the Local Group, and are 
described as the ``Council of Giants'' (CoG) by \cite{mccall_council_2014}; we adopt this CoG naming convention hereafter. 

\begin{figure}
\centering
\includegraphics[width=1.0\linewidth]{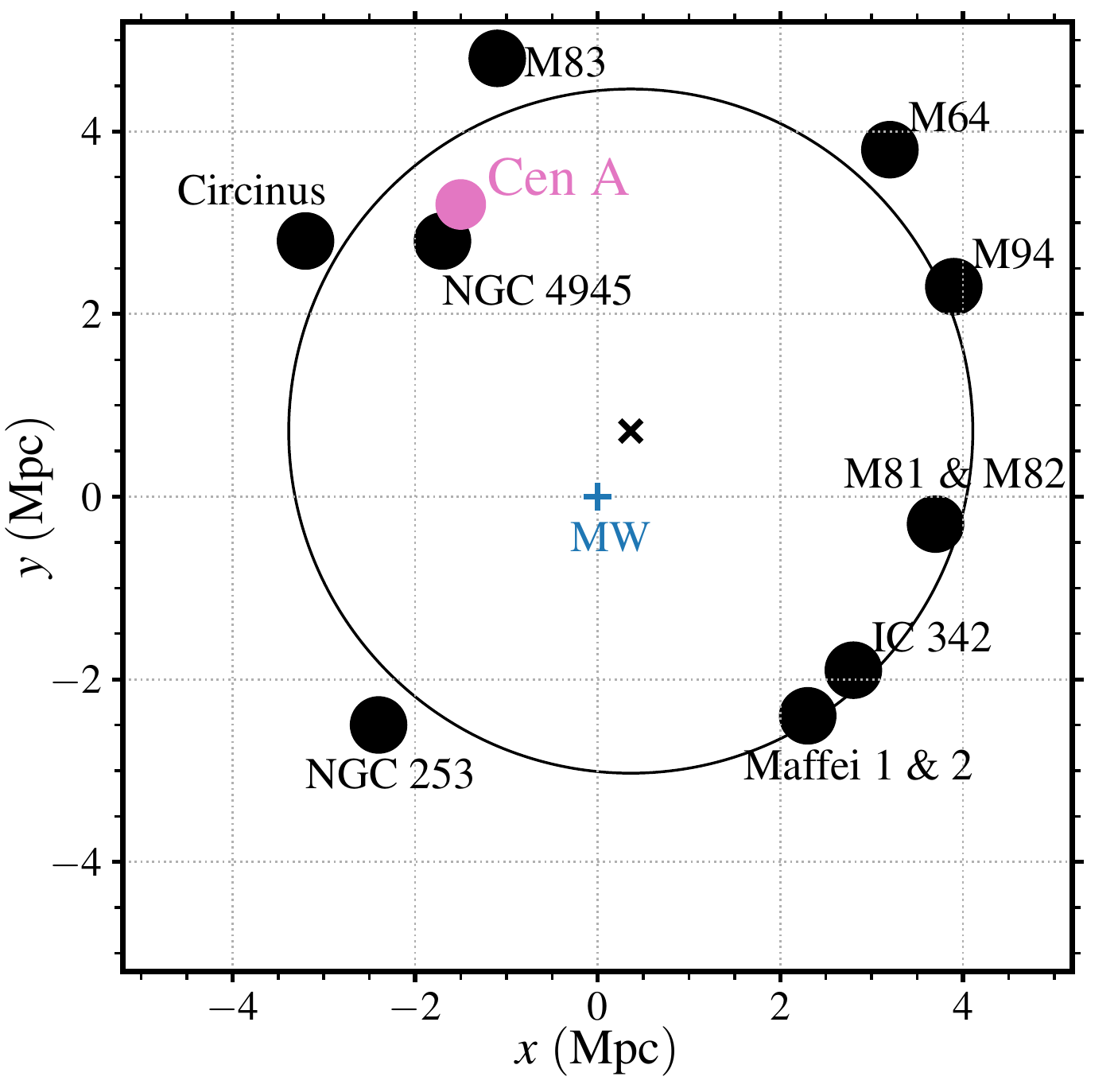}
\caption{The ``Council of Giants'' within the Local Sheet: A 2D diagram of the source (Cen~A; pink circle), observer (Milky Way; blue ``$+$'') and 9~scattering galaxies (black circles) used in this work. The solid black line marks a circle of radius $3.746$~Mpc and centred on $x=0.362~{\rm Mpc}$, $y=0.718~{\rm Mpc}$ (see ``$\times$'' in diagram), as defined by \protect\cite{mccall_council_2014}. The object positions in the diagram are provided in local sheet coordinates, in which the objects are predominantly located in the $x$-$y$ plane. 
}
\label{fig:CoG}
\end{figure}

The CoG or Local Sheet structure has a predominantly planar (ie. 2D) geometry and is approximately circular in structure. We consider here all members from the CoG listed by \cite{mccall_council_2014}. Fig.~\ref{fig:CoG} shows a depiction of the CoG objects which we focus on here in our study, shown in local sheet coordinates. The position of the MW is also indicated in this figure in blue, located close to the origin of the local sheet coordinate system. The position of the center of the best-fit circle describing the CoG members locations is indicated in Fig.~\ref{fig:CoG} as a black cross, and is located $\approx 0.8$~Mpc from the MW.

\begin{figure}
\centering
\includegraphics[width=1.0\linewidth]{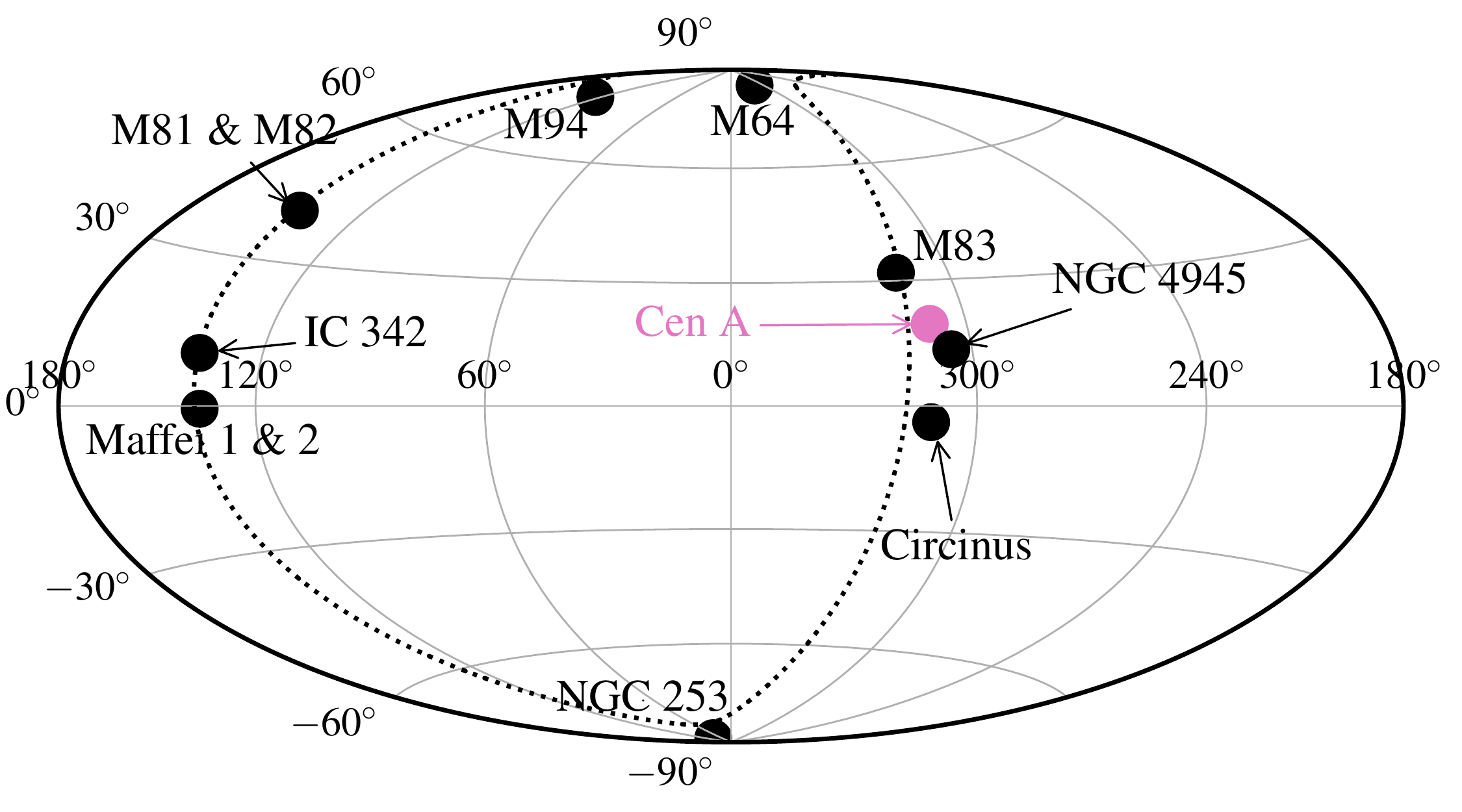}
\caption{A skymap showing the positions in the sky of the Council of Giant/Local Sheet objects. Cen~A is marked with a pink circle, the other council members are marked with black circles and the supergalactic plane is shown as a dotted line. 
}
\label{fig:CoG_skymap}
\end{figure}

The plane of these CoG objects, as observed by a terrestrial observer, is shown in Fig~\ref{fig:CoG_skymap}, in a Galactic coordinate representation (Hammer-Aitoff projection). In both Fig.s~\ref{fig:CoG} and \ref{fig:CoG_skymap}, the position of Cen~A within the CoG group is indicated in pink. The Galactic coordinates of, and distances to, the CoG members are given in Appendix~\ref{appendixa} together with stellar masses, estimated star formation rates (SFRs) and infra-red luminosities.

Within the CoG group, only Cen~A is known to demonstrate clear recent AGN jet activity, although Circinus may also exhibit some evidence of such activity, see \cite{1998MNRAS.297.1202E}. Definitive evidence for this activity in Cen~A is revealed by the radio emission from two giant inflated lobe structures extending out to $\approx 300~{\rm kpc}$, a distance scale comparable to the virial radius of its host galaxy \citep{1958AuJPh..11..400S}. In addition to this it also exhibits smaller inner lobes, indicating the onset of more recent AGN  activity \citep{croston2009}. 
Amongst the CoG members, no other objects display such prominent AGN jet activity, although the galaxies NGC~253 and M~82 do reveal heightened levels of star formation around their nuclear regions, with thermal X-ray images of these objects indicating the presence of outflow-like structures emanating from them \citep{1995ApJ...439..155B,Pietsch:2000cq}. 
Such outflows could potentially pollute the environment out to and beyond their virial radius with hot gas and magnetic field, as has been suggested to have occurred from recent analysis of a group of local galaxies (including NGC~253, M64, M81, M83, and M94) \citep{Bregman:2021kds}.

As well as affecting their environments powerful AGN and galactic outflows can accelerate particles to high energies. The maximum characteristic particle energy can be estimated from the Hillas-Lovelace condition, given by 
\begin{eqnarray}
E_{\rm max} \lesssim \frac{Z}{\eta}\left(\beta L_{\rm KE} \alpha \hbar \right)^{1/2}\approx  10~\frac{Z}{\eta}\left(\frac{\beta L_{\rm KE}}{3\times 10^{43}~{\rm erg~s}^{-1}}\right)^{1/2}~{\rm EeV}.
\label{Eqn:Hillas_Lovelace}
\end{eqnarray}
Here $\beta$ is velocity of the outflowing magnetised jet plasma in speed of light units, $L_{\rm KE}$ is the kinetic power, $Z$ is the atomic number, $\eta$ describes the scattering rate in units of the Bohm level scattering, $\alpha$ is the electromagnetic fine structure constant, and $\hbar=h/2\pi$ where $h$ is Plancks constant. This condition can be used to identify viable sites of UHECR acceleration. Energetically, the contents of the lobes of Cen~A are estimated to be $10^{59-60}$~erg, suggesting a time-averaged luminosity of $\sim 5\times 10^{43}$~erg~s$^{-1}$ assuming a slow subsonic inflation of the lobes \citep{wykes_mass_2013}. For an inflation velocity faster than this, the timescale for inflating the lobes is shorter, requiring a significantly higher jet power, potentially approaching the Eddington luminosity value. By comparison, an estimate of the kinetic luminosity of the winds of local starburst galaxies is more than an order of magnitude smaller than this \citep{heckman_nature_1990}. Furthermore, the velocities of these winds are themselves orders of magnitude smaller than AGN outflow velocities. The kinetic luminosity for Cen~A, and its outflow velocity, therefore indicates that it is unique amongst the CoG group as being the only member capable of satisfying the Hillas-Lovelace condition for particle acceleration to multi-EV rigidities (see eqn~\ref{Eqn:Hillas_Lovelace}). 

The thermal and magnetic pressures between galaxies within the CoG, and how the pressures at the center of the galaxies reduce with increasing distance from them, remains poorly understood. Observationally, there is a growing body of evidence that a ``warm–hot intergalactic medium'' (WHIM) permeates the space between galaxies \citep{2020Natur.581..391M}. Related, and conceptually similar, is the circumgalactic medium (CGM), usually defined as the region beyond the galactic disc but within the galactic virial radius \citep{tumlinson2017}, though it may in fact even extend beyond this radius \citep{2021ApJ...912....9W}. Collectively, the WHIM and CGM appear to account for a significant fraction of the "missing baryons" \citep{Gupta:2012rh,nicastro2018,Martynenko:2021aro}. Although low in density ($n\sim 10^{-(4-5)}~{\rm cm}^{-3}$), the high temperature of this gas ($kT\gtrsim 300$~eV) indicates that it provides significant thermal pressure within the extended galaxy out to distances comparable to the galactic virial radius ($\sim $300~kpc). Should the strengths of the magnetic fields, $B$, embedded within this gas be in approximate equipartition with the thermal energy density, $B^{2}/8\pi \approx nkT$ (i.e. the ratio of thermal to magnetic pressures, $\beta_{\rm B}$, is of order unity), magnetic field strengths within the range $0.05-0.2$~$\mu$G would also be expected out at these extended galactic distances. In reality, $\beta_{\rm B}$ in the CGM is not well constrained and could lie in the range $1-100$ \citep{pakmor2020,Faucher-Giguere:2023lgr}, likely varying within and between different objects. However, as discussed by \citetalias{bell_echoes_2022}, there is good evidence for large-scale magnetic fields surrounding M82, and our estimate of a $\sim 0.1{\rm \mu G}$ field at the virial radius is consistent with recent results from CGM modelling \citep{pakmor2020,vandevoort2021,2020ApJ...893...82F,faerman2023} and observation \citep{2023arXiv230206617H}, supporting the earlier suggestion that giant magnetised haloes (which we hereon refer to as \haloes) around nearby galaxies give rise to the anisotropy in UHECR skymaps \citepalias{bell_echoes_2022}.

As the CoG members possess a variety of SFRs (see Appendix~\ref{appendixa}), particularly in their Galactic nuclear regions, the levels of magnetisation of their \haloes ~at the virial radius are likely to vary considerably. However, given the current uncertainty on the physics dictating the driving of magnetic field and gas material to fill this region, for the sake of simplicity we here approximate that the \haloes ~of all CoG members to have the same sizes and magnetic field strengths; however, we discuss the possible hierarchy of circumgalactic magnetic field strengths and coherence lengths in CoG members (and their relative effectiveness as UHECR scatterers) further in section~\ref{sec:limitations}. 

\section{Simulation Setup}
\label{sim_setup}

To simulate the propagation of cosmic rays from Cen~A through the CoG structure to Earth, we adopt a Monte Carlo description. This description traces the spatial trajectory of the cosmic ray nuclei through the CoG system for a simulation timescale of 45~\Myr, with the first particles launched at $t=0$ such that the first particles arrive at Earth after $\approx 12~{\rm Myr}$. Since we remain agnostic about the actual location of the acceleration site, we adopt as generic as possible initial starting conditions for the simulation, assuming a uniform distribution of starting particles ($1.5\times 10^{9}$ in total) within a sphere of radius 300~kpc, centered on Cen~A. We adopt an isotropic distribution of initial particle momenta, which is motivated by the expectation that the particle Larmor radii are small compared to the dimensions of the jet dimensions on these spatial scales. The injected particles represent cosmic ray nuclei, with weighting factors being adopted so as to take into account their injected energy spectrum (see \ref{inj_spec} below). The positions of the scattering regions within our simulations are provided in Table~\ref{Table:CoG_Objects}. A planar view of the CoG system that the cosmic ray nuclei propagate through is provided in Fig.~\ref{fig:CoG}.

\subsection{Injected Energy Spectrum}
\label{inj_spec}
We inject particles into the system at Cen~A with a spectral energy distribution of
the form 
\begin{eqnarray}
\frac{dN}{dE}=\sum_{i=1}^{i_{\rm max}}f_{i}\left(\frac{E}{E_{0}}\right)^{-2} e^{-E/(Z_{i} R_{\rm max})}, 
\end{eqnarray}
where a spectral index of $2$ has been adopted, as motivated by Fermi diffusive shock acceleration theory for the case of strong shocks \citep{1994ApJS...90..561J}. In the above expression, the terms $f_{i}$ are the abundance of species of $i$, and $R_{\rm max}$ is the maximum rigidity that the UHECR source accelerates particles up to. A value for $E_{0}$, the minimum energy scales particles are injected at, of 30~EeV is adopted. This value for $E_{0}$ is adopted so as to focus our simulations on the energy scale at which small scale anisotropies are observed in the UHECR skymap data (see section~\ref{intro}). For our simulations, $R_{\rm max}=30$~EeV, which is compatible with the expectations found for scenarios in which the UHECR originate from a local source \citep{Taylor:2015rla,PierreAuger:2016use}. 

For the simulations considered here, a two species setup is adopted (ie. $i_{\rm max}=2$), consisting of \He\ and \Fe\ nuclei. Our choice of only a two component, light and heavy, mixed nuclear composition is simplistic, but deliberate. The fragility of the light He species above an energy of $10^{19.5}$~eV  \citep{hooper_intergalactic_2007,Wykes:2017nno}, with a loss length of just below 10~Mpc, motivates it as a natural diagnostic of the propagation time of the UHECR in the extragalactic radiation field environment. Likewise, the relative stability of the heavy Fe species, with a loss length of almost 3~Gpc at these energies, provides a contrasting reference population of particles with which to compare the light species abundance. In this description, the heavy species injected at the source can be considered as a crude proxy for species heavier than \He, and are considerably more stable than \He\ for the energies considered.

We adopt abundance ratios for He and Fe injected at the sources of $f_{\rm He}=0.868$, and $f_{\rm Fe}=0.132$, which, due to the impact of the cutoff already being felt at our adopted minimum energy, result in a ${\rm He}:{\rm Fe}$ ratio of $80:20$ at energy $E_{0}$. We adopt this ${\rm He}:{\rm Fe}$ ratio so as to give a comparable level of signal (within a factor of 3) in the Model C skymaps from both the direct and echoed waves. Our composition diagnostics are designed to be illustrative rather than providing a realistic match to UHECR composition as inferred from, e.g., $X_{\rm max}$ distributions \citep{PierreAuger:2014sui,PierreAuger:2014gko}; we reserve this exercise for future work.  

\subsection{Scattering Rates}
Our description adopts isotropic scattering rates for the interaction of the cosmic rays with the  magnetic fields local to the CoG objects. Due to a current poor knowledge of the magnetic fields on scales of the virial radius surrounding galactic structures, for simplicity we adopt an energy independent isotropic scattering rate for all cosmic ray nuclei in the system, with a scattering length,
\begin{equation}
l_{\rm sc}= 
\begin{cases}
    c\tau_{\rm sc} ,& \text{if } r\leq r_{\rm sc}\\
    \infty,              & \text{otherwise}
\end{cases}    
\end{equation}
where $r_{\rm sc}$ is the galactic scattering radius for the CoG members, which we fix to have a size of 300~kpc for all objects, a value close to the expected virial radii for a $10^{12}~M_{\odot}$ mass galaxy, and $r$ is the cosmic ray's distance from the CoG object. For our description of the scattering events, we allow for large angle isotropic scatterings to occur once the particles are within a distance $r_{\rm sc}$ of the scattering radius of each CoG galaxy. Our scattering description here differs in several ways from that adopted in \citetalias{bell_echoes_2022}. We assume large angle scattering from all CoG members. In contrast, \citetalias{bell_echoes_2022} adopted a small angle scattering description from only local objects with the largest SFRs. For our results here, a scattering time of $\tau_{\rm sc}=0.5$~Myr is adopted (ie. $r_{\rm sc}=150$~kpc). In comparison to this length scale, the Larmor radius of a 10~EV cosmic ray in a 0.1$\mu G$ magnetic field is $100$~kpc. Outside of the CoG galaxy regions, we assume that no scattering events take place.

\subsection{Photo-disintegration Rates}
We consider photo-disintegration of the cosmic rays in both the cosmic microwave background (CMB) and extragalactic background light (EBL) radiation fields. The photo-disintegration rates of UHECR nucei with the background radiation fields are determined by a convolution of the photo-disintegration cross-section with the radiation field spectral energy distribution \citep[][see their eq. 3]{hooper_intergalactic_2007}). For the photo-disintegration cross-section, we use the family of Lorentzian models proposed by \cite{Khan:2004nd}. For the EBL radiation field, we adopt the model from \cite{Franceschini:2008tp}.

\subsection{Coordinate System}
A coordinate system which aligns to that of the local sheet is adopted for the Monte Carlo simulations. The coordinate system can be related to Galactic coordinate system via the the rotation matrix,
$$
{\bf x}_{\rm Gal}={\bf M} ~{\bf x}_{\rm ls},
$$
where the rotation matrix ${\bf M}$ is given by

\begin{equation}
{\bf M}= 
\begin{pmatrix}
c_{\omega} s_{\delta} + s_{\omega} c_{\delta} s_{\alpha} & -c_{\omega} c_{\delta} + s_{\omega} s_{\delta} s_{\alpha} & -s_{\omega} c_{\alpha} \\
-s_{\omega}s_{\delta} + c_{\omega}c_{\delta} s_{\alpha}& s_{\omega}c_{\delta}+c_{\omega}s_{\delta} s_{\alpha}& -c_{\omega}c_{\alpha} \\
c_{\delta}c_{\alpha} & s_{\delta}c_{\alpha} & s_{\alpha} \\
\end{pmatrix}
\label{matrix}
\end{equation}

The angles for this rotation are $\alpha=172^{\circ}$, $\delta=225^{\circ}$ and $\omega=47.7^{\circ}$. Note the expression given in eqn~\ref{matrix}, utilise a shorthand notation in which $s_{\alpha}$ is used as an abbreviation for $\sin(\alpha)$ and $c_{\alpha}$ as an abbreviation for $\cos(\alpha)$.

\subsection{Cen~A Emission and Release Models}

In the following we describe the results for 3 different source evolution models. These models vary both the UHECR source luminosity evolution with time, and escape time of particles from the source region. We label these models A, B and C. The basic premise here is that model A allows us to explore the impact of the CoG structure on the UHECR signals, whereas models B and C can be considered representations of plausible physical scenarios. 

In {\bf Model A}, we consider the case in which the UHECR source (Cen~A) releases a single pulse of particles at $t=0$, with the particles subsequently escaping immediately from the source region. This model has a source term which is a $\delta$-function in time, whose resultant transmission through the system to an observer at Earth essentially provide a response or transfer function of the UHECR signal at Earth to the CoG structure. 

\begin{figure*}
\centering
\includegraphics[width=1.0\linewidth]{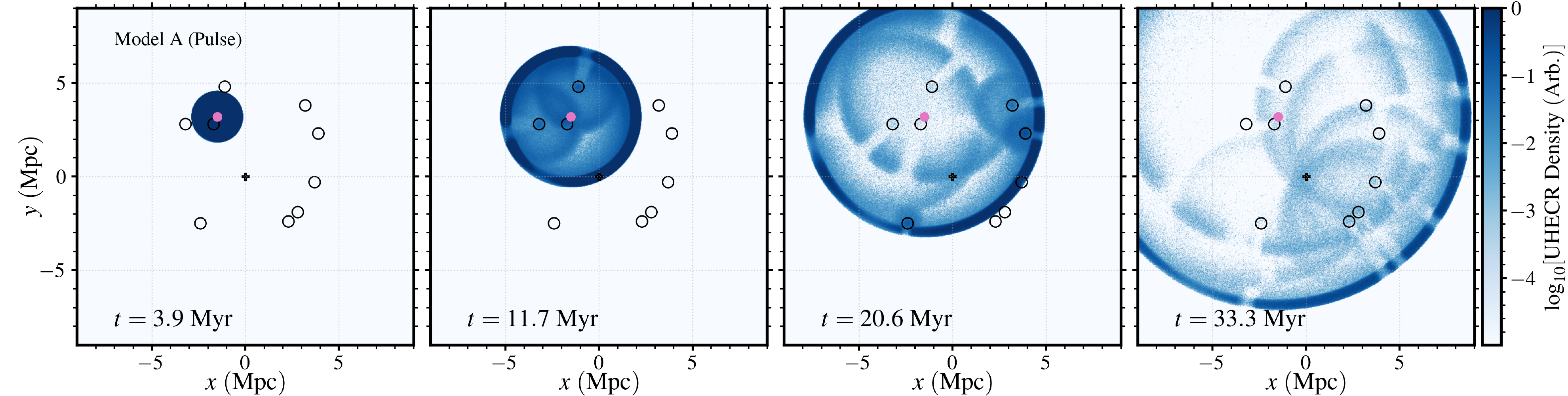}
\caption{Particle position maps from a slice of thickness $\Delta z = 0.6~{\rm Mpc}$ in the $z=0$ plane from Model A at four timesteps (3.9~\Myr, 11.7~\Myr, 20.6~\Myr, 33.3~\Myr), following their impulsive release from Cen~A. The corresponding plot for models B and C can be found in Appendix~\ref{appendixb}, and an animated version can be found in the online repository (see {\em Data Availability}). The position maps are presented as binned particle densities with bin sizes of $0.03~{\rm Mpc}$ and a density floor of $10^{-10}~{\rm bin}^{-1}$ in these arbitrary units. 
}
\label{fig:positions}
\end{figure*}

In {\bf Model B}, we consider the case in which Cen~A's UHECR source luminosity decreases exponentially over time after an initial outburst episode. For this model, once produced by the source, the particles escape immediately fron the source region. Using the timescale for the initial outburst as a reference time scale for our results ($t=0$), the subsequent UHECR luminosity is given by,
\begin{eqnarray}
L=L_{0}e^{-t/{\tau}_{\rm dec}},\ ({\rm for}~t>0)
\end{eqnarray}
where $\tau_{\rm dec}$ is the decay time of the UHECR source luminosity, which we set to 3~Myr. Such a short activity timescale would be consistent with AGN flickering model proposed to describe the activity evolution of other local AGN \citep{2009BASI...37...63S}. However, such a description is a crude approximation to the true variability in the UHECR luminosity of Cen~A. We apply an additional Gaussian smoothing with standard deviation of $1$~\Myr\ to the launch times of the CR particles. The purpose of this smoothing --  which is applied to both Models B and C, but not Model A --  is to limit the sensitivity of our results to the exact timestamps we present, which is an appropriate choice given the uncertain time evolution of Cen~A. By incorporating this smoothing, we ensure that the results presented would be insensitive to any details of initial conditions on smaller spatial scales than the injection radius.

In {\bf Model C}, we consider the case in which Cen~A injects a pulse of particles, with the particles subsequently residing longer within the source region. As was done for Model~B, particles are injected with a Gaussian distribution of times centred on $t=0$, with a standard deviation of $1$~\Myr\ (i.e. Gaussian smoothing in the time domain was carried out). Contrary to the rigidity independent description we adopt for particle propagation through the CoG structure, we approximate the physics of diffusive escape out of the magnetised lobes of Cen~A by imposing an additional rigidity-dependent escape time 
for each particle, given by
\begin{equation}
    \tau_{\rm esc} = \tau_{10} 
    \left( 
    \frac{E/Z}{10~{\rm EV}}
    \right)^{-1},
\label{eq:escape}
\end{equation}
where $E$ is the particle energy, $Z$ the particle charge, and $\tau_{10}$ is the escape time for a $10~{\rm EV}$ rigidity particle, for which we choose $\tau_{10}=1.5~{\rm Myr}$. Such an escape time for particles with rigidity 10~EV is consistent with these particles experiencing around 1 scattering event before being able to escape from their host environment region. While $t<\tau_{\rm esc}$, a given particle can undergo photo-disintegration loss interaction, but it does not move from it's starting position; only after $t=\tau_{\rm esc}$ does the particle start propagating. Although this description fails to capture any change in rigidity of the particle as it undergoes energy losses, such rigidity changes during photo-disintegration are minor.

\section{Results}
\label{results}

\subsection{Particle Spatial Distribution}

Following the propagation of cosmic ray nuclei through the CoG system, the arrival of multiple waves of particles to the MW location are observed. Fig.\ref{fig:positions} shows a $z=0, ~\Delta z=0.6~{\rm Mpc}$ slice of the particle spatial distribution in the system for four key timescales: 0~\Myr, 12~\Myr, 21~\Myr, and 33~\Myr, in the $x-y$ (local sheet) plane. Fig.~\ref{fig:positions} shows a snapshot of the logarithm (base 10) of the binned density in the simulation (bin size 0.03 Mpc) at four different times during the simulation foe Model~A. Also shown in this figure are the positions of the CoG objects (empty circles), Cen~A (pink filled circle), and the MW location (black vertical cross). From the snapshots of the particle density at the four timescales, the arrival of waves to the MW location at 12~\Myr, 21~\Myr, and 33~\Myr\ can be seen.

\subsection{Direct and Echoed Waves}

\begin{figure}
\centering
\includegraphics[width=\linewidth]{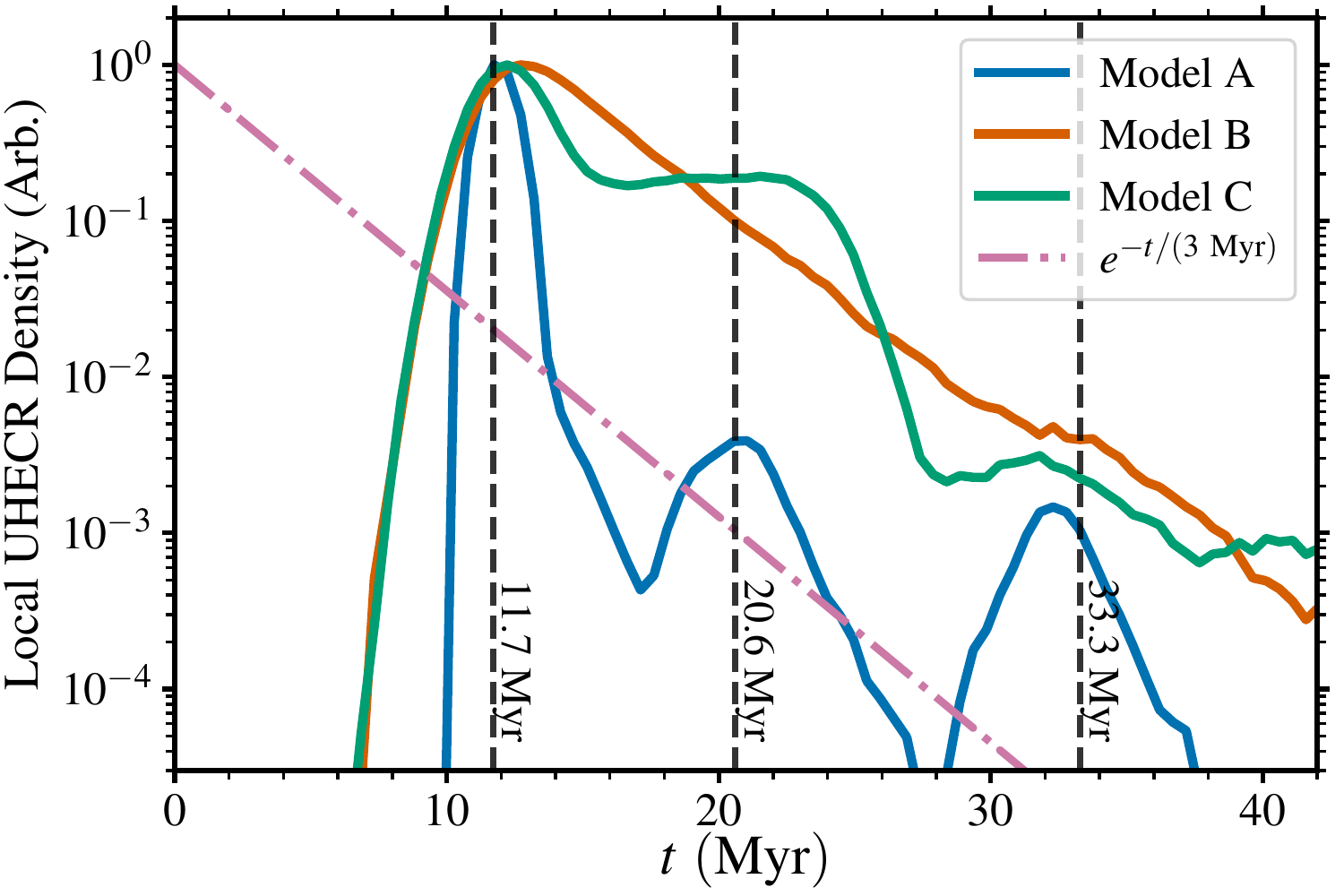}
\caption{The CR particle density in a local box of size 300~kpc, centered on the Milky Way location, following the injection of particles from Cen~A. The three colours show results from the three models considered: the single pulse (blue), declining source (orange) and ridigity-dependent escape (green). The red dotted line shows an exponential with a decay time of $3$~\Myr, and 
the dashed vertical lines mark $t=11.7,20.6,33.3$~\Myr, the times at which $x-y$ positions in Figs~\ref{fig:positions} and skymaps in Figs~\ref{fig:skymap_a},\ref{fig:skymap_b}, \ref{fig:skymap_c}  are shown. 
}
\label{fig:waves}
\end{figure}

The four key timescales noted, relate to the intitial spatial distribution (0~Myr), the arrival of the direct wave from Cen~A (12~\Myr), and the arrival of two echoed waves (21~\Myr~ and 33~\Myr). The arrival of the direct and echoed waves can be easily appreciated from the blue line in Fig.~\ref{fig:waves}, which shows the arriving UHECR density as a function of time after the initial outburst from Cen~A. Three major peaks are observed in this figure, namely the direct wave at 12~\Myr\ after the initial outburst, and the echoed waves at 21~\Myr\ and 33~\Myr\ after the outburst. The width of the peaks of the waves seen in the figure result from the finite size of the scattering regions.

An understanding of the different timescales which these waves arrive from, and the specific sources responsible for contributing to the echo signal, can be understood from Fig.~\ref{fig:ellipse}. This figure provides a dissection of the echoed waves, connecting their contribution to sources located on a common concentric ellipse, whose two focii are located at Cen~A and the MW. The colour scale in the figure indicates the incurred delay time for each concentric ellipse.

\subsection{Focusing Effects}
\label{sec:focusing}
\begin{figure}
\centering
\includegraphics[width=\linewidth]{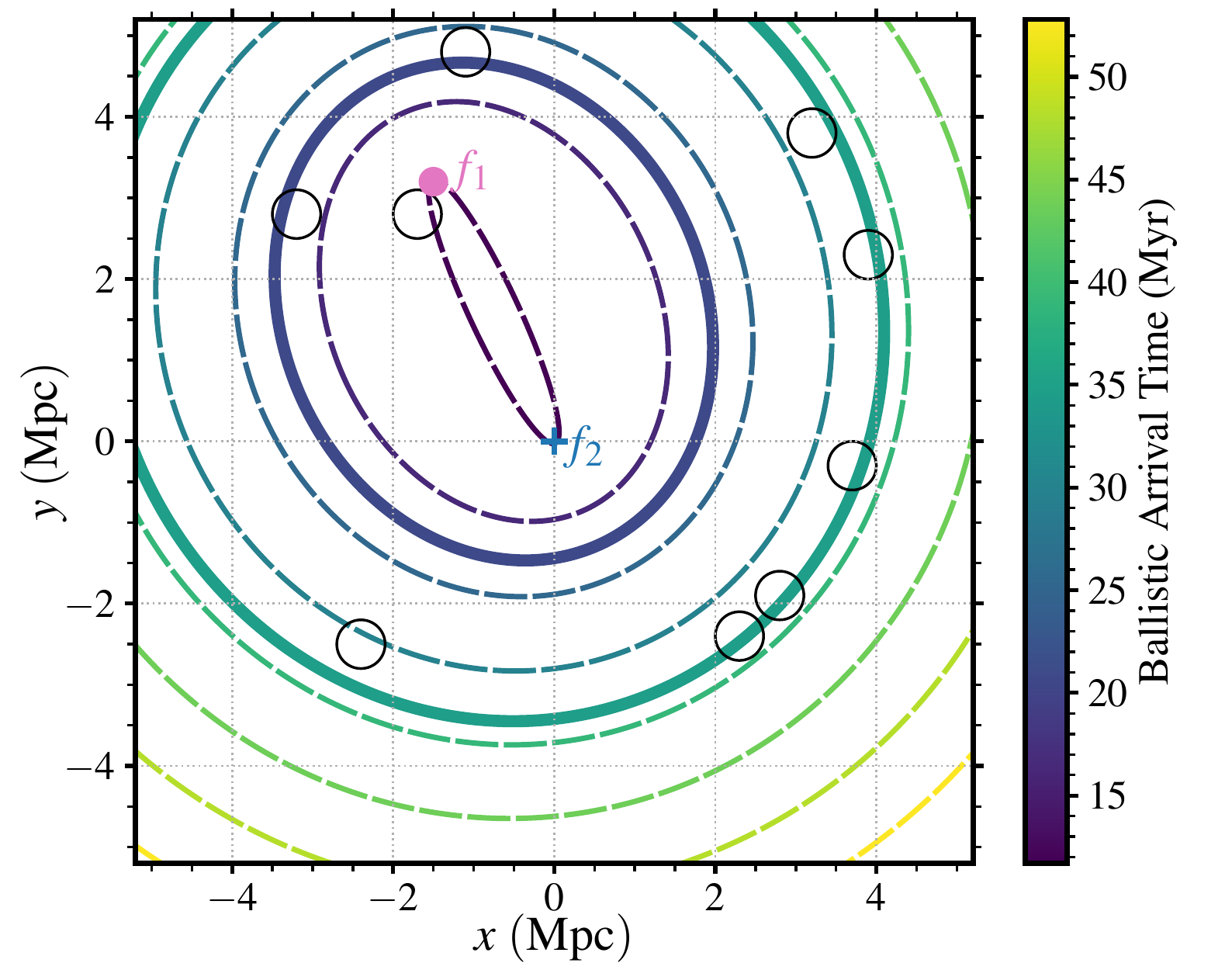}
\caption{
A family of `isodelay contours', which form concentric ellipses with a variety of eccentricity values, colour-coded by the ballistic time of arrival. The ellipses are plotted as dashed lines from $11.7$ to $52.8$~\Myr\ at $2.94$~\Myr\ intervals, with additional thick solid lines overlaid for $t=20.6$~\Myr\ and $t=33.3$~\Myr\ (see also Fig.~\ref{fig:waves}. The two focal points ($f_{1}$ and $f_{2}$) of the ellipses are centered on Cen~A and the Milky Way, respectively, and the positions of the council of giants are marked with open circles. The relationship between the ballistic arrival time and eccentricity is given in the text, with larger eccentricities $e$ corresponding to earlier arrivals.
}
\label{fig:ellipse}
\end{figure}

As appreciated directly from the particle density snapshots shown Fig.~\ref{fig:positions}, the scattering of particles from the CoG objects results in the arrival to the MW of focussed waves of particles. To understand the origin of this focussing effect, Fig.~\ref{fig:ellipse} shows a family of concentric ellipses (of varying eccentricity), with each ellipse having Cen~A and the MW at the two focal points. These curves represent isotemporal contours for signals from Cen~A which arrive to the MW at the same time. As observed in this figure, the CoG objects are approximately located on specific concentric ellipses (blue and yellow thick solid lines in Fig.~\ref{fig:ellipse}), where the colour of the ellipse indicates the corresponding delay time incurred.

The eccentricity, $e$, of an ellipse with the source and observer at the two focii can be related to the (straight-line, ballistic) time of arrival as $e = R / (ct)$. Here $R$ is the distance to the CR source, and $t$ is the time of arrival of scattered CRs with respect to the initial burst. With these definitions, the first CRs arrive at $t \approx R/c \approx 12~$\Myr. Subsequent echoes from sources lying on an ellipse with eccentricity $e$ arrive at $t \approx R/(ce)$, with a delay with respect to the light travel time of $\tau \approx R (1-e) / (ce)$. This geometrical argument assumes ballistic trajectories in between scatterers. Any additional small angle scattering introduced during particle propagation in the IGM would also introduce  delays, though such delays would be expected to be safely negligible if the IGM scattering angle is itself small.

Fig.~\ref{fig:ellipse} shows that the two echo waves after the initial outburst are caused by the collective influence of CoG members that are located approximately on the $t \approx 20$~\Myr\ and $t \approx 33$~\Myr\ isodelay contours. Specifically, the first echo wave is caused by scattering by M 83 and Circinus, and the second wave is caused by six sources located approximately on the same isodelay contour. One particular source, NGC~4945 is also responsible for a number of interesting effects in our simulations, due in part to its close proximity to Cen~A. NGC~4945 intercepts rather a large fraction of the CRs coming from the source, some of which are scattered towards Earth, enhancing the signal from the approximate Cen~A direction slightly and spreading out the arrival times. There is also a shielding effect from NGC~4945, which happens to lie on an approximately straight line path between Cen~A and NGC~253. CRs are attenuated and fewer CRs reach NGC~253, which acts to weaken the NGC~253 echo signal in the resulting skymaps (see section~\ref{sec:skymaps}).

\subsection{Local Skymaps}
\label{sec:skymaps}

\begin{figure}
\centering
\includegraphics[width=1.0\linewidth]{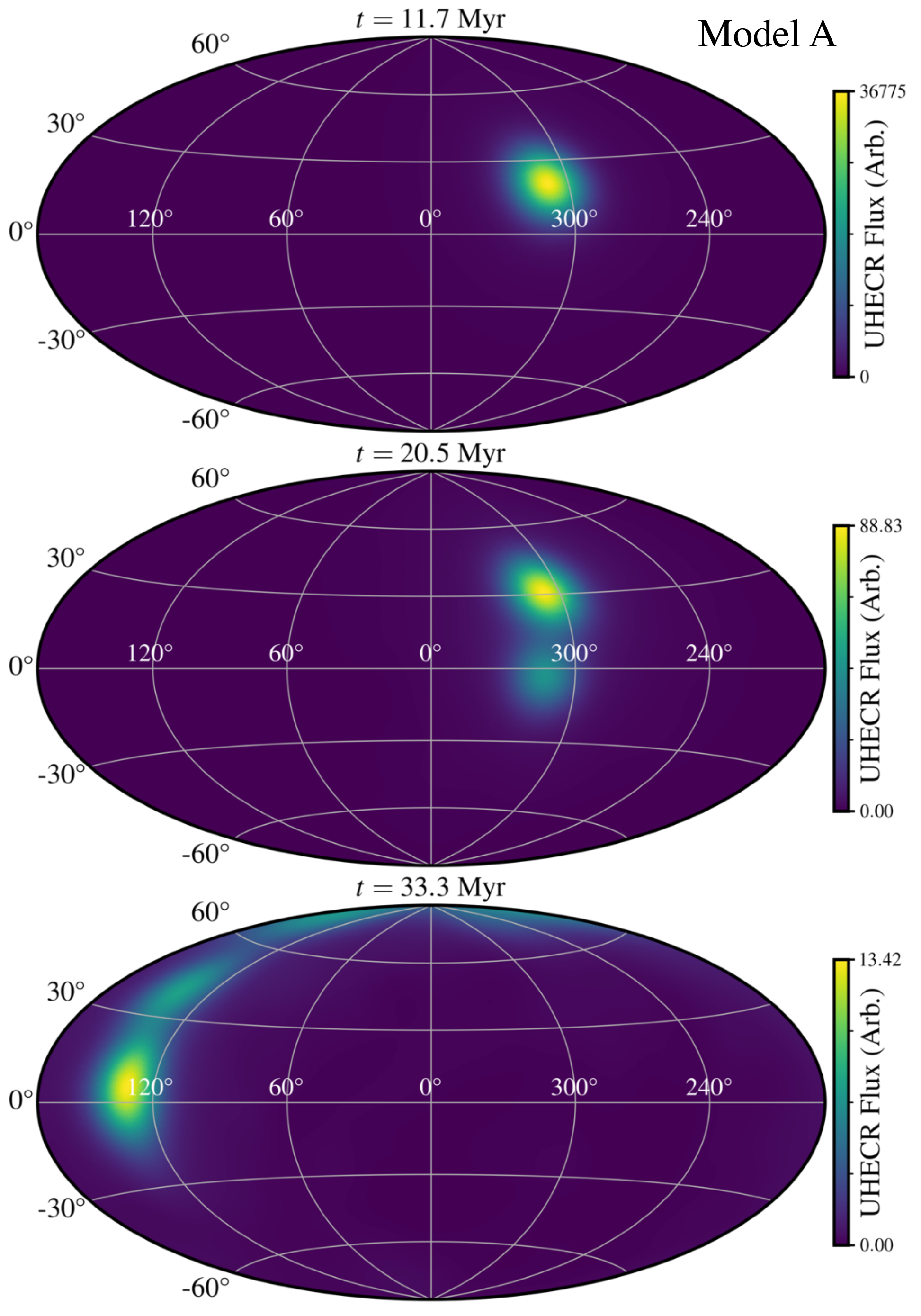}
\caption{Three skymaps in Galactic coordinates (Hammer-Aitoff projection) from Model~A at $11.7$~\Myr\ (top), $20.6$~\Myr\ (middle), and $33.3$~\Myr\ (bottom) after the impulsive cosmic ray release from Cen~A. The colour-scale encodes the number of particles per HEALpix pixel, initially calculated with $32\times32$ pixels covering the sky, which has then been smoothed with a Gaussian symmetric beam with full-width at half-maximum of $20^\circ$. Animations of all skymaps are available in an online repository (see {\em Data Availability}).}
\label{fig:skymap_a}
\end{figure}

The angular distribution of particles arriving to an observer located in the MW (i.e. the arriving particle skymap), after scattering from the CoG objects, is shown in Galactic coordinates in Fig.~\ref{fig:skymap_a} for the model~A scenario. The different panels in this figure show arriving cosmic ray skymaps at 11.7, 20.6, and 33.3~\Myr\ after a Cen~A outburst of UHECR. To produce these skymaps, we binned the arrival directions into solid angle bins, in Galactic coordinates, using the Healpy python implementation \citep{zonca_healpy_2019} of the HEALpix scheme \citep{gorski_healpix_2005}. The colour-scale in these skymaps encodes the number of particles per HEALpix pixel (ie. solid angle bin), initially calculated with $64\times64$ pixels covering the sky. 

In contrast to \citetalias{bell_echoes_2022}, we do not include small-angle scattering in the regions between galactic \haloes. Instead, to approximate this process, we apply a Gaussian angular dispersion to the particles contributing to the skymaps. The value for the angular dispersion level adopted is motivated by considering the angular deflection of particles with a Larmor radius, $r_{\rm Lar}$, propagating a path of length $d$ through an ensemble of coherent magnetic field patches, each of size $l_{\rm coh}$. In such a case, the root mean square deflection angle is
\begin{eqnarray}
\delta\theta &=& \frac{l_{\rm coh}}{r_{\rm Lar}}\left(\frac{d}{l_{\rm coh}}\right)^{1/2} \nonumber\\
&\approx& 2^{\circ}~Z \left(\frac{l_{\rm coh}}{1~{\rm Mpc}}\right)^{1/2}\left(\frac{d}{4~{\rm Mpc}}\right)^{1/2}\left(\frac{B}{0.5~{\rm nG}}\right)\left(\frac{30~{\rm EeV}}{E}\right)
\end{eqnarray}
where $r_{\rm Lar}=E/ZeB$ is the Larmor radius. 
We therefore apply a rigidity-dependent Gaussian dispersion of $(E/30~{\rm EeV})^{-1} Z\times 2^\circ$ on a particle by particle basis. Such an approach is of course only an approximation of the amount of IGM scattering as a function of sky position, but is broadly appropriate given that the source and scatterers are all located at a similar distance from Earth; nevertheless, we discuss this limitation further in section~\ref{sec:limitations}. Additionally, all our skymaps are smoothed with a $20^{\circ}$ Gaussian smoothing, so that they can be directly compared with the most recent PAO results \citep{PierreAuger:2022axr}, for which a comparable level of smoothing is adopted. We note that the angular size of this Gaussian smoothing is larger than the $\approx 5^\circ$ angular radius subtended by a $300~{\rm kpc}$ \halo ~at $3.7~{\rm Mpc}$.

As expected from the spatial distribution results discussed above, the early time direct wave (12~\Myr) originates from Cen~A. The arrival of the first echo wave to the MW at 21~\Myr, originates from the CoG objects close to Cen~A (NGC~4945, M83 and Circinus), as expected from the delay time ellipses shown in Fig.~\ref{fig:ellipse}. In contrast to these two earlier skymaps, the arrival of the second echo wave to the MW at 33~\Myr, originates from CoG objects located further from Cen~A, on the side opposite to the location where Cen~A resides. The results in Fig.~\ref{fig:skymap_a} show how the CoG structure reverberates to a pulse of CRs, and can thus be thought of as a sparse representation of a spatially-resolved response function, analogously to the transfer and response functions used in spectroscopic reverberation mapping of AGN \citep[e.g.][]{blandford1982,peterson1993}. The observed signal is then a convolution of the results from Model A with the underlying activity evolution of the source.

\begin{figure}
\centering
\includegraphics[width=1.0\linewidth]{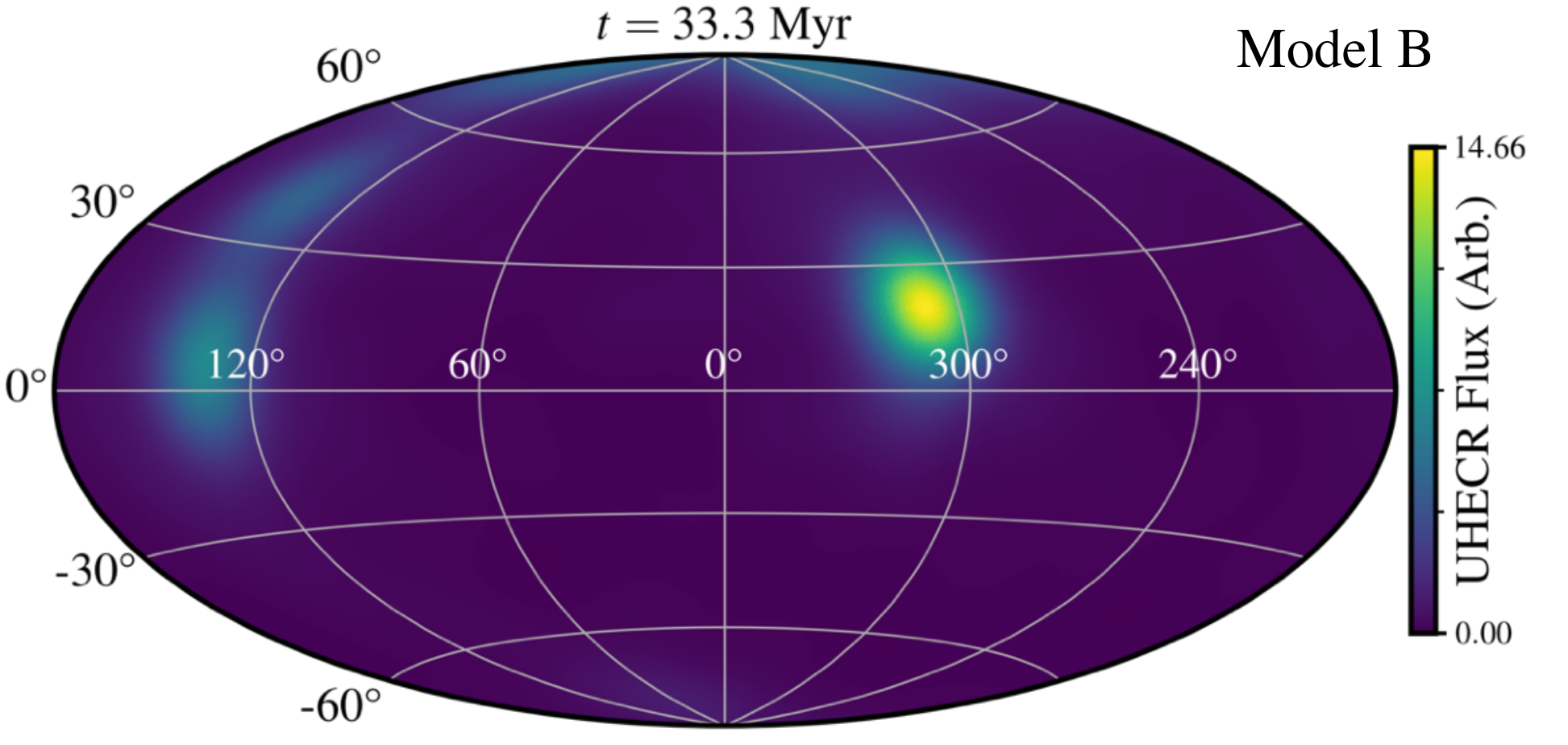}
\caption{Skymap in Galactic coordinates (Hammer-Aitoff projection) at $33.3$~\Myr, for Model~B, the declining source scenario, for which a decay time of $\tau_{\rm dec}=3$~Myr has been adopted. The map is calculated in the same way as in Fig.~\ref{fig:skymap_b}. 
}
\label{fig:skymap_b}
\end{figure}

\begin{figure}
\centering
\includegraphics[width=1.0\linewidth]{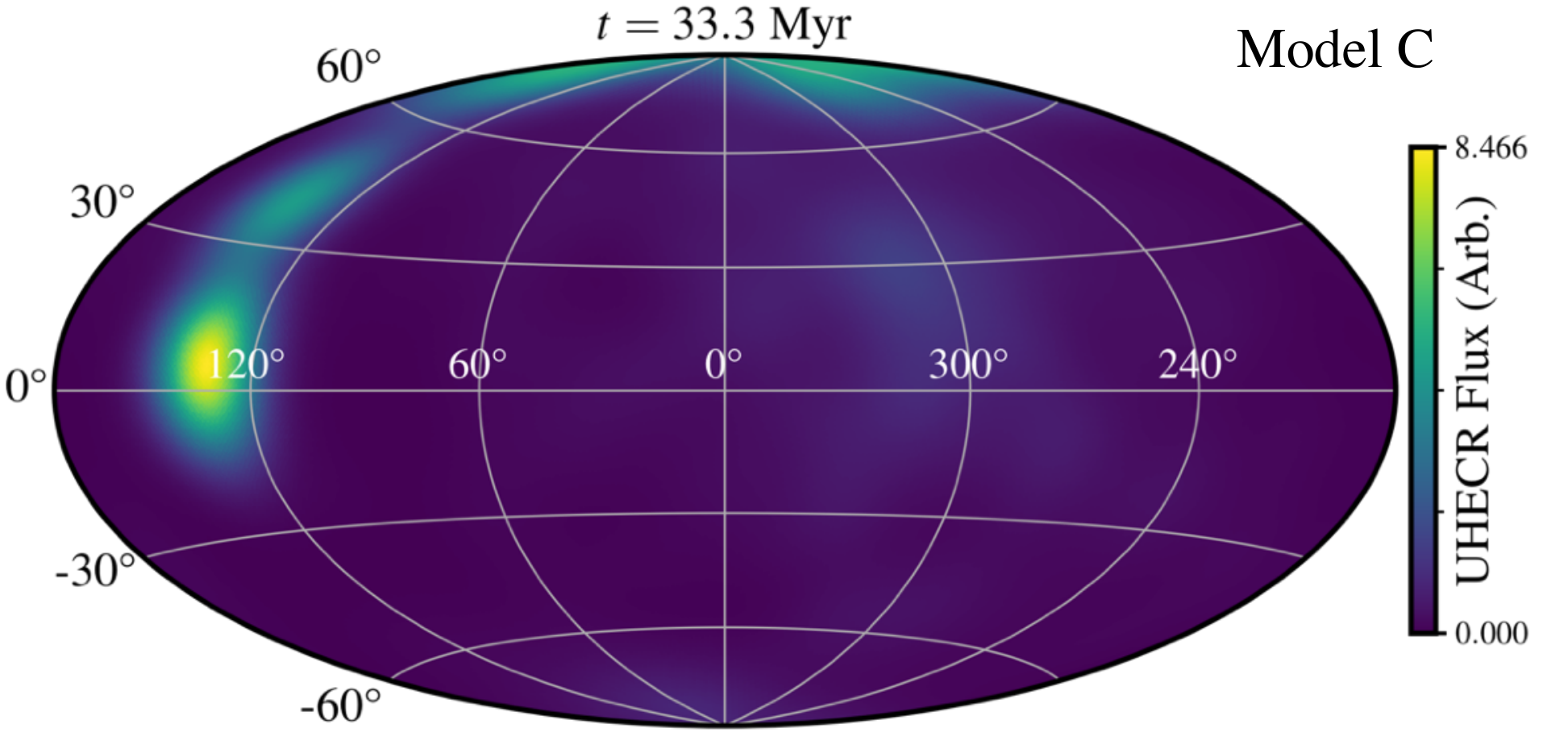}
\caption{As Fig.~\ref{fig:skymap_b}, but for Model~C, the delayed escape scenario in which particles have a rigidity-dependent escape time described by equation~\ref{eq:escape} with $\tau_{10} = 1.5~{\rm Myr}$.
}
\label{fig:skymap_c}
\end{figure}

Similar plots are shown in Fig.s~\ref{fig:skymap_b} and \ref{fig:skymap_c} for model~B and model~C outburst scenarios, respectively, focusing now only on the $t=33.3$~\Myr\ snapshot. In our framework, and as also suggested by \citetalias{bell_echoes_2022}, we consider this time period to be a reasonable approximation to the present day in the sense that it represents a characteristic time elapsed since Cen~A was at its peak of UHECR activity. At earlier times in these simulations the snapshots only have small variations in the anisotropy and are dominated by signal from the Cen~A direction (as can be seen from the online animations). However,  the arriving UHECR flux at late times (33~Myr) allows for bright spots of comparable intensity in the skymap for both direction towards Cen~A, and towards the CoG members located furthest from Cen~A. We note that the hotspot observed from the direction of Cen~A in the model~C skymap is, however, significantly more extended and diffuse than the hotspot observed from Cen~A in the model~B skymap. This increase in the hotspot extension for model~C is due to the dominance of heavy species in the Cen~A signal for this case (see section~\ref{composition_skymaps}). These skymaps, for both model~B and model~C, show striking similarities with the observational results from both the PAO and TA \citep{TelescopeArray:2014tsd,2018ApJ...853L..29A,PierreAuger:2022axr}, in particular when compared to the all-sky anisotropy patterns \citep{biteau2019,dimatteo2020}. Specifically, a signal is observed from the direction of Cen~A, with a ring of additional hotspots produced by echoes from the directions of Maffei/IC~342, M81/M82, M94 and M64. The relative brightness of the Cen A signal and the echo signals depends on the model parameters -- in particular, the adopted composition, the parameters controlling the source activity ($\tau_{\rm dec}$) and subsequent CR escape ($\tau_{\rm esc}$), the angular dispersion $\delta \theta$, and the halo sizes $r_{\rm sc}$. Different relative intensities and extensions can be achieved by tuning these parameters accordingly.

\begin{figure}
\centering
\includegraphics[width=1.0\linewidth]{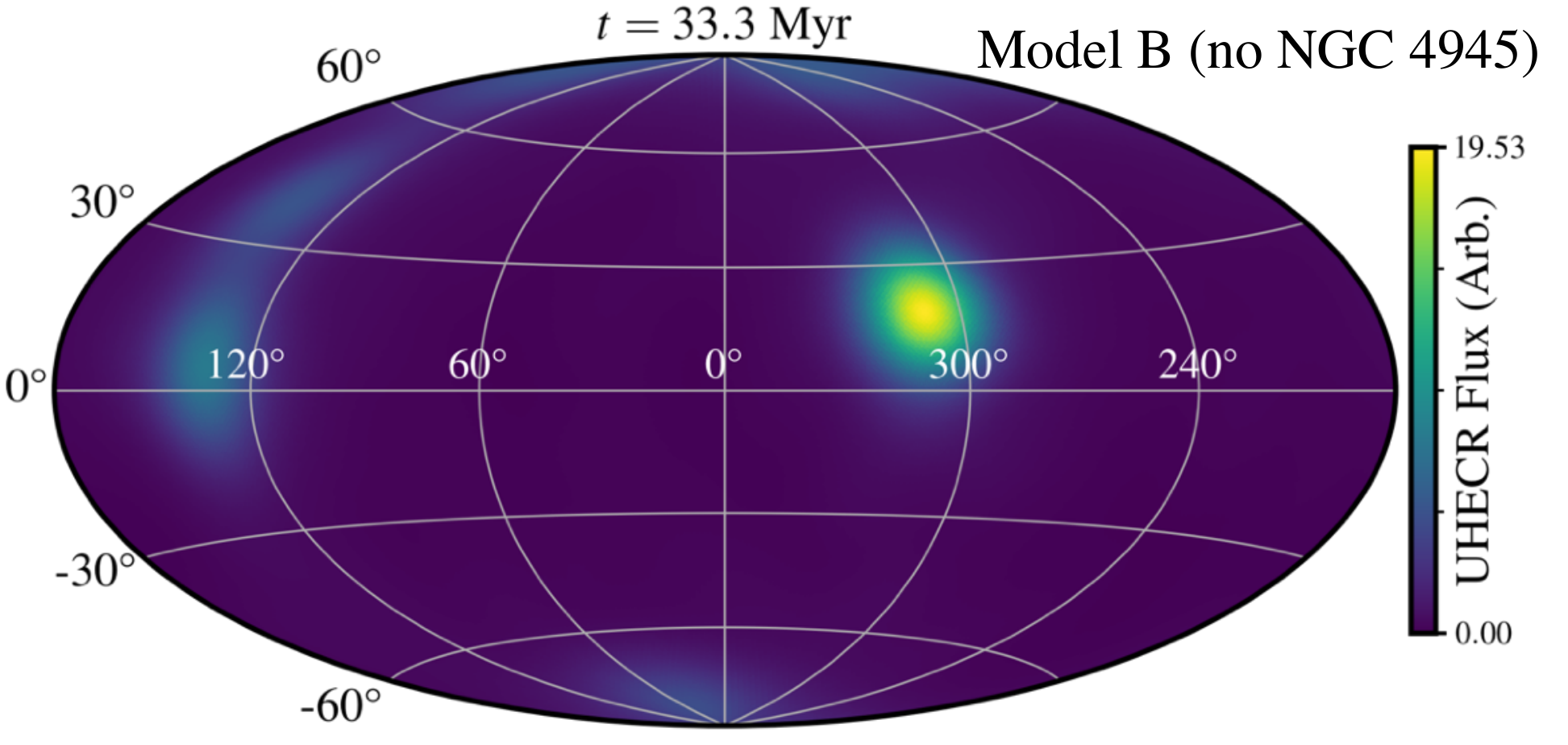}
\caption{As Fig.~\ref{fig:skymap_b}, but for Model~B with NGC 4945 removed from the simulation. Removing the shielding impact of NGC~4945 results in a stronger excess in the direction of NGC~253 at southern Galactic latitudes as discussed further in the text. 
}
\label{fig:skymap_4945}
\end{figure}

It is worth commenting on the conspicuous absence of NGC~253 from the late-time skymaps. As noted above in section~\ref{sec:focusing}, NGC~4945 creates a shielding effect that significantly decreases the UHECR flux impinging on the NGC~253 \halo; this effect is responsible for the negligible signal at late times from the direction of NGC~253. To demonstrate this, in Fig.~\ref{fig:skymap_4945}, we present results from a simulation identical to Model~B, but with NGC~4945 removed. In this modified simulation a hotspot is indeed produced from the direction of NGC~253 at southern Galactic latitudes. There is a fairly prominent excess in this region of the sky in the PAO maps, so for the echo model to explain this we would either require some variation between the \haloes' ability to scatter UHECRs (as might be expected anyway; see section~\ref{sec:limitations}), or for additional scattering in the IGM to allow UHECRs to be deflected around NGC~4945. Alternatively, an additional local source near to the Galactic south pole could contribute, such as the Fornax A radio galaxy \citep{matthews_fornax_2018,eichmann_ultra-high-energy_2018}. 

Our results from both Models B and C are fairly similar to those presented by \citetalias{bell_echoes_2022}, who show skymaps in equatorial coordinates, with a few differences. \citetalias{bell_echoes_2022} focused mainly on the TA hotspot and the influence of the M82 galaxy, before presenting a simulation which included M82, NGC~253 and IC~342. Our results show that this qualitative match to the observed skymaps does not disappear when photo-disintegration losses are included, as would be expected for the relatively short propagation times. Furthermore, we have included additional sources and so observe additional hotspots in the direction of M94 and M64, while Maffei 1\&2 act to smear out and enhance the feature near IC~342. Finally, we note the influence of NGC~4945, Circinus and M83. These sources are close to Cen A on the sky and, depending on the model and timestamp, can act to produce a smeared out or elongated pattern in the direction of Cen~A. In particular, in some cases the `Cen~A' feature resembles a lop-sided dumb-bell shape, better correlated with M83. This is an interesting general point given that the hotspot observed from PAO is somewhat diffuse (a top-hat search radius of $\approx 25^\circ$ is found by \citealt{PierreAuger:2022axr}), and not perfectly aligned with Cen~A \citep{aab_indication_2018}; we therefore suggest that scatterers local to the source may be important in determining the morphology of any observed excesses.

\subsection{Composition-dependent Skymaps}
\label{composition_skymaps}

In order to obtain further insights into the results show in Fig.s~\ref{fig:skymap_b} and \ref{fig:skymap_c}, it is helpful to consider a breakdown of these results into the contributions from different logarithmic nuclear species (ie. $\ln A$) groups. 

\begin{figure*}
\centering
\includegraphics[width=1.0\linewidth]{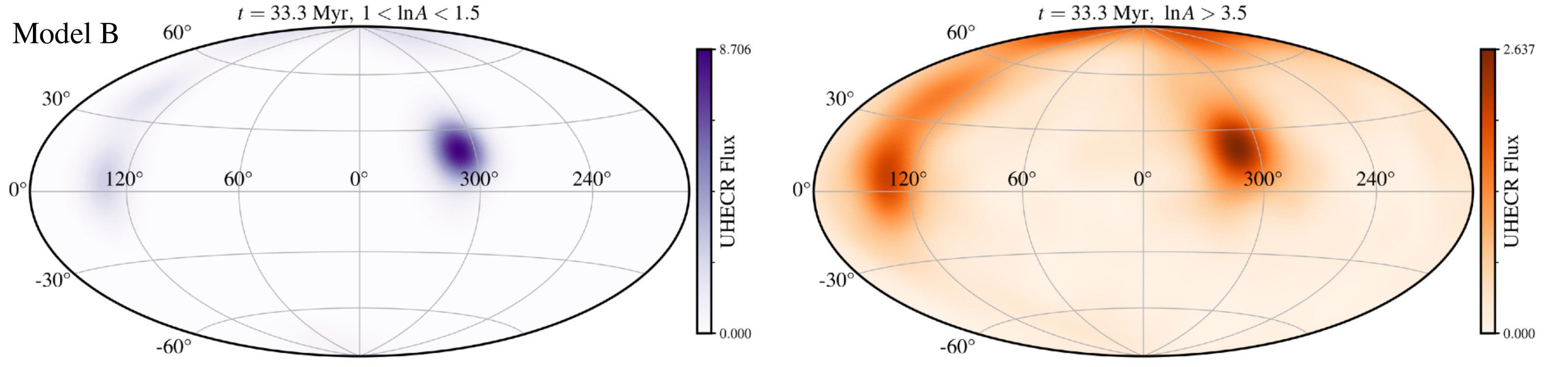}
\caption{Composition-dependent skymap in Galactic coordinates (Hammer-Aitoff projection) for Model B, the declining source scenario. The left-hand panel (purple) shows the results from $1 < \ln A < 1.5$, spanning the \He\ mass range ($A = 3-4$), and the right-hand panel (orange) shows the results from $\ln A > 3.5$, spanning the \Fe\ mass range ($A = 34-56$). As in Fig.~\ref{fig:skymap_a}, the plots are constructed using the HEALpix scheme and a Gaussian smoothing function of $20^\circ$.
 In this model, `echo' features from the CoG members at large angular distances from Cen~A are only significant in the higher mass bin ($\ln A > 3.5$), and the low-mass bin is dominated by relatively He-rich CRs that were accelerated more recently by Cen~A. 
}
\label{fig:skymap_comp_b}
\end{figure*}

In Fig.~\ref{fig:skymap_comp_b} such a decomposition of the model~B skymaps in Fig.~\ref{fig:skymap_b} into composition groups is shown for the mass ranges $1<\ln A<1.5$, and $\ln A > 3.5$. As is seen from this figure, for model~B the \He\ component ($1<\ln A<1.5$) of the arriving flux from CoG group members furthest from Cen~A is considerably depleted relative to the \He\ component from Cen~A. Contrary to this, the \Fe\ signal contribution ($\ln A > 3.5$) from these two regions in the sky are similar in magnitude.

\begin{figure*}
\centering
\includegraphics[width=1.0\linewidth]{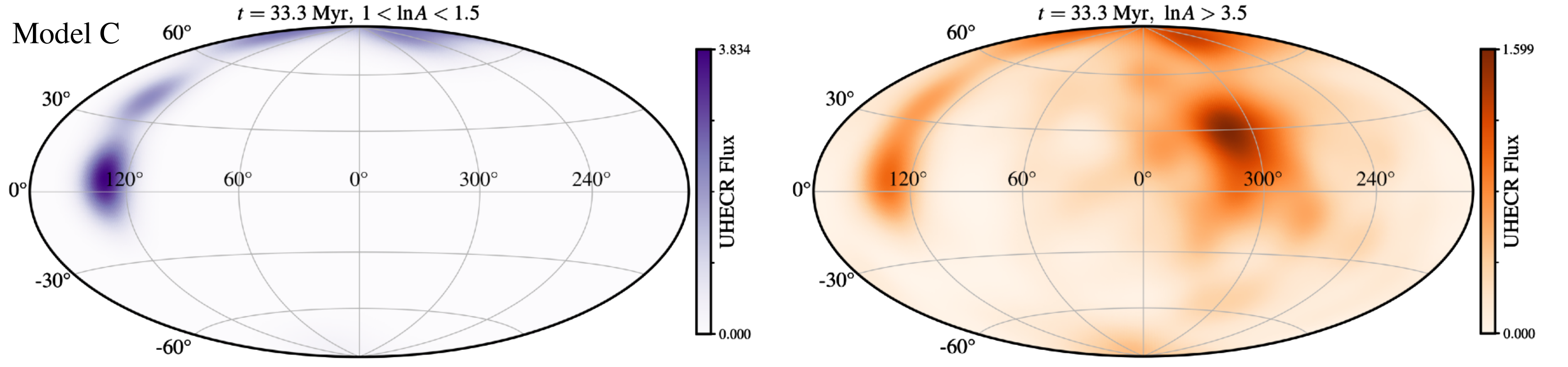}
\caption{As Fig.~\ref{fig:skymap_comp_b}, but for Model C, the delayed escape scenario. In this model, `echo' features from the CoG members at large angular distances from Cen~A are only significant in the lighter mass bin ($1 < \ln A < 1.5$), and the high mass bin is dominated by Fe-rich CRs that have escaped Cen~A more recently and scattered off NGC~4945, M83 and Circinus. 
}
\label{fig:skymap_comp_c}
\end{figure*}

In Fig.~\ref{fig:skymap_comp_c} a decomposition of the model~C skymaps in Fig.~\ref{fig:skymap_c} into composition groups is shown for the mass ranges $1<\ln A<1.5$, and $\ln A > 3.5$. This figure shows that for the model~C case, the signal from Cen~A at late times is almost purely \Fe\ dominated. As noted earlier in section~\ref{inj_spec}, it should be borne in mind here that the \Fe\ species in these results should be considered as a proxy of species heavier than \He. In contrast, the signal at late times from the CoG objects furthest from Cen~A is almost purely \He\ dominated in our simulations. From Fig.s~\ref{fig:skymap_comp_b} and \ref{fig:skymap_comp_c}, it is also generally apparent that, as expected from the imposed ridigity-dependent dispersion, the hotspot regions associated in the heavy species skymaps are considerably more extended than the hotspot regions observed in the light species skymaps. We have not explored this effect in detail in this paper, but the angular dispersion of different composition signals represents another diagnostic that merits future study.

We note that we have repeated these simulations for both model~B and C, with the \Fe\ nuclei injected at the source instead replaced by \Ox\ nuclei, which has lower $Z$ and a shorter energy loss length (although both are still considerably larger than for \He). For model B, this test reveals that apart from the reduced level of angular dispersion of the particles in the skymaps (see section~\ref{sec:skymaps}) and some small changes in the relative intensity of the direct and echoes signals, the results are largely unchanged. For model C, similar results can also be achieved, although the degree of similarity depends on whether $\tau_{10}$ is kept fixed or is scaled accordingly. This sensitivity on $\tau_{10}$ comes about because the its value was chosen to give roughly comparable intensity from Cen~A and the echo signal at $\approx 33$~Myr. Thus, while the results inevitably change at a quantitative level when different compositions are adopted, our tests act as a verification that the \Fe\ nuclei in our simulations can indeed be considered as an illustration of the general behaviour of a species heavier than \He.


These composition-dependent skymap results for both model B and C demonstrate the insight, in addition to  the usual angular information, that can be provided by the composition information. UHECR nuclei, operating as "composition clocks", can provide the additional third dimension to skymaps, giving rise to a clear skymap signature for a particular propagation scenario from a local source. 

\section{Discussion}
\label{discussion}

This work builds further on the possibility that UHECR at the highest energy may have a local extragalactic origin \citep[][\citetalias{bell_echoes_2022}]{Wykes:2017nno}. Such a possibility appears compatible with the evidence both that a local UHECR source must exist \citep{Taylor:2011ta,lang_revisiting_2020}, and that a small number of such sources are contributing to the UHECR flux observed at Earth \citep{Ehlert:2022jmy}. We now discuss our results within the wider astrophysical context, focusing on the key uncertainties in our model, before exploring the prospects for testing the echoes model in the future.

\subsection{Scattering in Local Extragalactic Magnetic Fields}
\label{sec:limitations}
One of the key aspects of our work is that the magnetised CGM of galaxies within the CoG must represent an effective barrier to UHECRs if they are to produce UHECR echoes. As discussed in section~\ref{intro} and by \citetalias{bell_echoes_2022}, while the magnetic fields in the CGM are uncertain, the field strengths required to deflect UHECRs are plausible. In our work, we made the simplifying assumption that each CoG member has the same optical depth and \halo ~size. Provided that the optical depth is larger than $1$ the results presented here are not found to be qualitatively sensitive to the specific value adopted. However, in detail neither of these assumptions is likely to hold, even if the pressures and densities in the respective CGMs are comparable. In particular, there is likely to be variation in the plasma beta value, $\beta_{\rm B}$, since magnetic fields can be amplified and stretched by dynamos and dynamical interactions, or transported from the galaxy to the CGM through outflows. It is important to note that the SFR within the Galactic nuclear regions of the COG members varies considerably. Assuming that the level of this central SFR activity dictates a galaxy's ability to drive material out into its \halo ~region, a large variety of \halo magnetic field strengths would also be expected. Subsequently, CoG members possessing the largest nuclear SFR would be expected to possess the largest optical depths.

In addition, the structure of the magnetic field is important, because there must be some ordering of the field on the scale of the UHECR Larmor radius. A discussion of the ability of M82 to produce large-scale, ordered magnetic fields is given by \citetalias{bell_echoes_2022}, but we also draw attention to the results of \cite{pakmor2020} who examine the magnetised CGM in spiral, MW-like, galaxies with ``zoom-in'' cosmological MHD simulations. They find the CGM is magnetised by an {\sl in situ} turbulent dynamo, which can create a magnetic field of strength $\sim 0.1~{\rm \mu G}$ by $z=0$. However, they also show that large-scale ordered fields are only produced in the presence of strong galactic outflows.

Taking all the above evidence together, the likely variation of the strength and structure of the circumgalactic magnetic field between galaxies would be expected to naturally create a {\em hierarchy}: some galaxies, perhaps those undergoing interactions or that have recently undergone a burst of star formation, would be effective UHECRs scatterers, while others could be more or less transparent to UHECRs. Such a hierarchy is likely to be important for explaining the apparent correlation of UHECR arrival directions with star-forming or starburst galaxies \citep{2018ApJ...853L..29A,PierreAuger:2022axr}, and possibly even necessary for explaining why the MW is not opaque to UHECRs (see discussion in section~\ref{MW_prop}).

Finally, we note the additional simplifying assumptions we have made. We neglected particle scattering in extragalactic space beyond the virial radius of the CoG objects (although we did approximate this effect by smoothing the skymaps), and within the virial radius of the MW (see discussion in section~\ref{MW_prop}). Additionally the contribution to the UHECR skymap from more distant sources have been neglected, and the scattering in the \haloes ~was treated independently of rigidity. Each of these assumptions is warranted of further interrogation, which we leave to future work.

\subsection{Propagation Within the Milky Way's Magnetic Field}
\label{MW_prop}

Given that only $\sim10\%$ of the UHECR detected by the PAO above 40~EeV appear anisotropic \citep{2018ApJ...853L..29A}, correlating with the CoG structure \citep{van_vliet_extragalactic_2022}, the results presented in section~\ref{results} are specifically focused on accounting for the origin of this anisotropic component. The origin of the remaining quasi-isotropic signal has been intentionally neglected. 


One possible way in which a level of quasi-isotropic signal could also be accounted for, expanding on the setup adopted in this study, would be the inclusion of particle scattering within the MW's own virial radius. The main effect expected from UHECR propagation within an extended, turbulent MW halo is an additional dispersion of the arriving anisotropic UHECR signal \citep{Shaw:2022lqd}. With the size of this dispersion being rigidity dependent, the spreading of the heavier species, for particles in a given energy band, could give rise a quasi-isotropic component in the skymap. In contrast, lighter species, whose angular dispersion would be smaller, would produce more anisotropic components in the skymap. A similar effect could also be induced by sufficiently strong ($\gtrsim {\rm nG}$) intergalactic magnetic fields, as hinted at by the greater angular spread in the high mass skymap in the right-hand panel of Fig.~\ref{fig:skymap_comp_c}.

Additionally, at sufficiently low energies, UHECR diffusion within the MW from an external source would be expected to give rise to a skymap with a largely dipolar anisotropy component \citep{Giacinti:2011uj}. Whether local propagation at these lower energies ($>8$~EeV) could account for the recent the discovery by the PAO of a dipole of magnitude $7\%$ in the UHECR skymap \citep{PierreAuger:2017pzq}, consistent with the weaker evidence for a dipole also seen by TA \citep{TelescopeArray:2020cbq}, remains an open question. With the magnitude of this dipole increasing with energy, potentially reaching a magnitude of more than 10\% above an energy of 40~EeV \citep{PierreAuger:2020fbi}, a possible connection to the scattering scenario we put forward here seems warranted for future investigation.

\subsection{Cen~A's Activity Evolution}
\label{Cen_A_evolution}
A key astrophysical aspect of the echoes scenario is that the original UHECR source must be variable, which is necessary for any of the echo waves or hotspots to be of comparable significance to the direct flux. More specifically, the source -- in our case, Cen~A -- needs to have declined significantly in UHECR luminosity over a $\sim 20$~\Myr\ timescale if the hotspots from the echoes are to be approximately the same intensity as the hotspot from the region of Cen~A. This decline can be achieved by a direct corresponding change in source power, or a combination of a change in power and the CR spectral index. In our modelling, the ratio of UHECR luminosity $20$~\Myr\ ago to  the present day luminosity is $\approx 700$. This factor, however, scales inversely with the cross-sectional area of the haloes (ie. with the square of the \halo ~radius). Increasing the scattering halo radius to 800~kpc, one of the values considered by \citetalias{bell_echoes_2022} and motivated by \cite{wilde2021,2020ApJ...900....9L}, would decrease this required magnitude of UHECR variability in Cen A to $\approx 100$. The larger radius, and a consequently stronger echo, might be appropriate for galaxies such as M82 which are more strongly star-forming.

As discussed previously in other papers \citep[][\citetalias{bell_echoes_2022}]{matthews_fornax_2018,matthews_ultrahigh_2019},  evidence exists supporting the possibility that Cen~A possessed enhanced activity in its `recent' history. 
Specifically, the inferred age of the synchrotron-emitting electrons in the giant radio lobes is $\sim 20-30$~\Myr\ \citep{hardcastle_high-energy_2009}, is comparable to the timescales of the echo waves considered here. Furthermore, Cen~A's giant lobes have an estimated total energy content of $\sim10^{59-60}~{\rm erg~s}^{-1}$ \citep{wykes_mass_2013,eilek_dynamic_2014}. If inflated over a similar $\sim 20$~\Myr\ timescale this would require a mean jet power of $\sim 10^{44-45}~{\rm erg~s}^{-1}$, some 1.5-15\% of Cen~A's potential Eddington luminosity, $7\times10^{45}~{\rm erg~s}^{-1}$, assuming a black hole mass of $5.5\times10^{7}~M_\odot$ \citep{cappellari_mass_2009}. These luminosity estimates for Cen~A are consistent with the required kinetic energy luminosity necessary for the source to be considered capable of accelerating UHECRs, as discussed in section~\ref{local_structure} (see Eqn.~\ref{Eqn:Hillas_Lovelace}).

\subsection{Predictions and Outlook}
\label{predictions}

The hotspot maps obtained from our simulations shown in Fig.s~\ref{fig:skymap_comp_b} and \ref{fig:skymap_comp_c} can be compared with the full-sky joint PAO/TA hotspot map, combining the data from PAO above 40~EeV and from TA above 53~EeV \citep[][see their Fig.~4, which also highlights the alignment of both the local sheet and supergalactic planes to these hotspots]{PierreAuger:2020mkh}. As apparent from a comparison of these simulation maps with the observational map, consistency between them can be found for either the model B and model C simulation scenarios. 

The coming advent of AugerPrime \citep{2019EPJWC.21006002C} provides an exciting test bed for looking for further insights in the anisotropy signature reported by the PAO \citep{2018ApJ...853L..29A,PierreAuger:2022axr} and TA \citep{TelescopeArray:2014tsd}. Observations by AugerPrime are anticipated to allow the composition of air showers to be probed on a shower-by-shower basis. We consider here how composition dependent skymaps will allow the model explored here to be tested.

Although models B and C lead to apparently similar skymaps at a time period of around 20~Myr after the Cen~A outburst event (see Fig.s~\ref{fig:skymap_b} and \ref{fig:skymap_c}), the composition dependent skymaps for this same time window shown in Fig.s~\ref{fig:skymap_comp_b} and \ref{fig:skymap_comp_c} are noticeably different. 

One general feature found is that the \He-like flux for the Cen~A region, and the CoG region away from Cen~A are strongly different. In model B, the \He\ flux for the echo signal from the region away from Cen~A is small compared to the \He\ flux for the Cen~A region. In contrast to this, for model C, this the \He\ signal from the region away from Cen~A is large compared to the \He\ signal from the Cen~A region.

The geometrical nature of the delayed signal from Cen~A, produced by UHECR echoes off the CoG structure,  predicts a similar level of brightness of UHECR signal from sources on the same isodelay contour (see Fig.~\ref{fig:ellipse}), as appreciated by Fig.~\ref{fig:skymap_b} and Fig.~\ref{fig:skymap_c}. Furthermore, the composition of the signal echoed off objects on the same isodelay contours should also match, as appreciated from Fig.~\ref{fig:skymap_comp_b} and Fig.~\ref{fig:skymap_comp_c}. The approximate equal brightness of the sources seen in these results, however, partly comes from the assumption of equal size scattering regions, $r_{\rm sc}$, for all CoG members. This assumption was made on the basis of simplicity rather than from observational motivations. In contrast to this dependence on underlying assumptions, the expectation of equal composition from objects on the same isodelay contour is a more robust prediction, being insensitive to the scattering region size.

Aspects of our findings here are more general than the specific Cen~A source scenario that we consider. Provided that the primary UHECR source resides sufficiently close, the composition of the direct and echoed waves of UHECR, following their release from the source, offer a key diagnostic to probing the probing both the location of the UHECR source, and the local magnetic environment.

\section{Conclusions}
\label{conclusion}

We here explore a potential origin of the observed correlation of UHECR with nearby extragalactic structures reported by PAO above an energy of 40~EeV \citep{2018ApJ...853L..29A,PierreAuger:2022axr} and TA above an energy of 50~EeV \citep{TelescopeArray:2014tsd}. Specifically, we investigate whether such a correlation can result from the echo signal of UHECR, originally accelerated and released by Cen~A, off the local extragalactic structure, developing further a scenario initially considered by \cite{bell_echoes_2022}. 

Focussing our attention on the CoG structure, the dominant extragalactic structure at distances $<10$~Mpc from the MW, we consider ballistic propagation of UHECR beyond 300~kpc from members of the CoG structure, with the UHECR undergoing large angle scattering on approaching distances smaller than this from any of the member objects. We find that the propagation of a pulse of UHECR from Cen~A through this structure gives rise to three distinct signals. The first signal at 12~Myr, is produced by direct wave from Cen~A. The second and third signals are the two echo waves at 21~Myr and 33~Myr.

Beyond these pulse results, we additionally consider the effect introduced by both Cen~A's activity evolution over the last 30~Myr (model B), and the rigidity dependence of the UHECR escape from Cen~A (model C). In both model B and C cases, it is shown that under reasonable assumptions for these two processes, hotspots corresponding to the CoG members in the late time ($> 30$~Myr) skymap are obtained, following the initial outburst from Cen~A (see Fig.s~\ref{fig:skymap_b} and \ref{fig:skymap_c}).

Through the consideration of the propagation of \He\ and \Fe\ nuclear species in the UHECR signal, and the photo-disintegration of these species en-route, we obtain composition dependent skymaps. These skymaps are produced by a mixture of direct and echoed signals. It is demonstrated that the apparent degeneracy in the late time skymaps for model B and C can be broken using the spatial distribution of the light component regions (see Fig.s~\ref{fig:skymap_comp_b} and \ref{fig:skymap_comp_c}). Furthermore, the echo origin of the CoG objects correlation, quite generally, predicts a common signal composition from all CoG members located on a common isodelay contour (see Fig.~\ref{fig:ellipse}). 

Our results suggest that the use of ``composition clocks'' -- that is, UHECR composition as a measure of the travel time for the UHECRs as a function of arrival direction and/or energy --- has more general and exciting prospects as a probe of the UHECR time domain, with the potential for testing the UHECR echo model as well as other UHECR source scenarios.

\section*{acknowledgments}
AT acknowledges support from DESY (Zeuthen, Germany), a member of the Helmholtz Association HGF. JHM acknowledges funding from the Royal Society, and, previously, from the Herchel Smith Fund at the University of Cambridge. ARB acknowledges the support of an Emeritus Fellowship from the Leverhulme Trust. We would like to thank Alan Watson, Foteini Oikonomou, Yakov Faerman and Arjen van Vliet for helpful discussions. This work was performed using resources provided by the Cambridge Service for Data Driven Discovery (CSD3) operated by the University of Cambridge Research Computing Service (\url{www.csd3.cam.ac.uk}), provided by Dell EMC and Intel using Tier-2 funding from the Engineering and Physical Sciences Research Council (capital grant EP/T022159/1), and DiRAC funding from the Science and Technology Facilities Council (\url{www.dirac.ac.uk}). We gratefully acknowledge the use of the following software packages: healpy \citep{zonca_healpy_2019}, astropy \citep{astropy_collaboration_astropy_2013,astropy_collaboration_astropy_2018}, matplotlib \citep{hunter2007}.

\section*{Data availability}
\label{availability}
Data and accompanying scripts to reproduce Figures 1 to 5 in this paper, together with animations of all skymaps and Fig.~\ref{fig:positions}, are available in a github repository (\url{https://github.com/jhmatthews/uhecr-echo-vis}) with an associated Zenodo DOI: \href{https://doi.org/10.5281/zenodo.7634625}{10.5281/zenodo.7634625}. The additional raw data to reproduce the skymaps are available from the authors upon request.   

\bibliographystyle{mnras}

\bibliography{bibliography}

\appendix
\section{Table of galaxy properties}
\label{appendixa}

In Table~\ref{Table:CoG_Objects}, we show the complete list of the CoG objects included in our calculations, together with their positions, stellar masses, infra-red luminosities, radio fluxes, and estimated SFRs. References for the sources of these estimates and measurements are given in the table caption, as are the symbol definitions. 

\begin{table*}
\centering 
\renewcommand{\arraystretch}{1.2}
\begin{tabular}{c c c c c c c }
  \hline 
  Galaxy & $l~(^\circ)$ & $b~(^\circ)$ & Distance~(Mpc) & $M_{*}$~($10^{10}~M_{\odot}$) & $L_{12\mu {\rm m}}$ ($10^{9}L_{\odot}$) & est. SFR ($M_{\odot}~{\rm yr}^{-1}$) \\
  \hline \hline 
  NGC~253   & $97.36$ & $-87.96$ & 3.5    & 1.7 & 3.5 & 5.4   \\
  M64       & $315.68$ & $84.42$ & 5.0    & 11.5 & 1.3 & 2.3   \\
  M81       & $142.09$ & $40.91$ & 3.7    & 7.1 & 0.4 & 0.8    \\
  M82       & $141.41$ & $40.57$ & 3.5    & 1.3 & 7.8 & 10.7   \\
  M83       & $314.58$ & $31.97$ & 4.9    & 2.7 & 3.4 & 5.2  \\
  M94       & $123.36$ & $76.01$ & 4.5    & 3.8 & 0.9 & 1.6   \\
  NGC~4945  & $305.27$ & $13.34$ & 3.3    & 1.2 & 1.8 & 3.0  \\
  IC~342    & $138.17$ & $10.58$ & 3.4    & 2.7 & 2.1 & 3.5  \\
  Maffei~1  & $135.86$ & $-0.55$ & 3.3    & 6.2 & -- & --   \\
  Maffei~2  & $136.50$ & $-0.33$ & 3.4    & 1.2 & 0.9 & 1.5   \\
  Circinus  & $311.33$ & $-3.81$ & 4.3    & 1.5 & 6.2 & 8.8   \\
  \hline &  \\
\end{tabular}
\caption{
Object names, Galactic coordinates ($l,b$), distances, stellar masses ($M_*$), $12~{\rm \mu m}$ lumnosity ($L_{12\mu {\rm m}}$), estimated SFRs of the CoG members included as UHECR scatterers in our simulations. Distances are taken from \protect\cite[][table~1]{mccall_council_2014} (Table~1). Mass in stars, infrared luminosities, and estimated SFRs of objects are taken from the WISE catalogue for extended sources \protect\citep{2019ApJS..245...25J}.}

\label{Table:CoG_Objects}
\end{table*}

\section{Particle Position Maps for Model~B and Model~C Cases}
\label{appendixb}

\begin{figure*}
\centering
\includegraphics[width=1.0\linewidth]{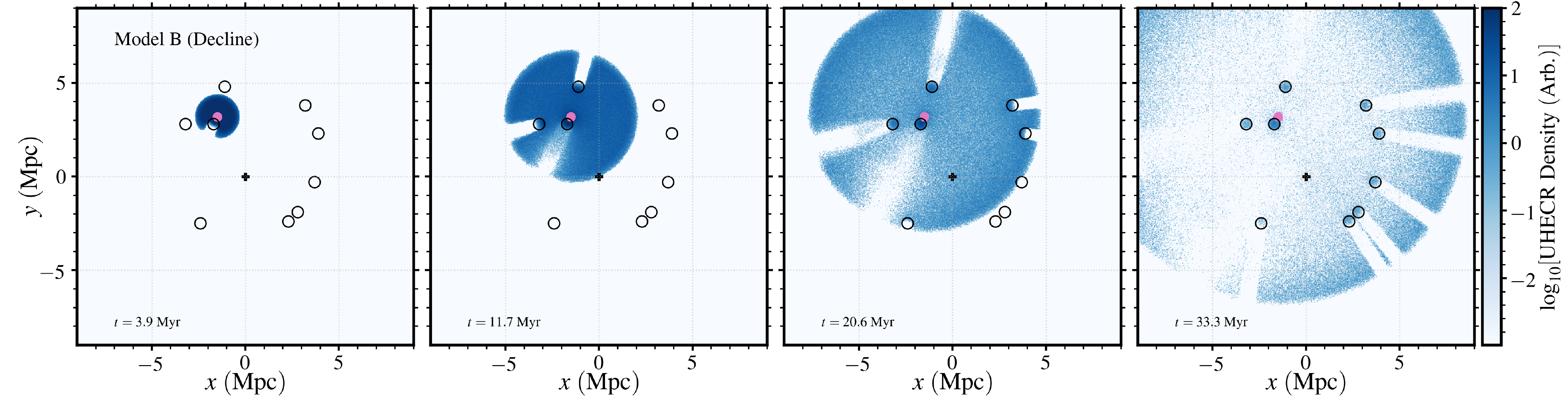}\\
\includegraphics[width=1.0\linewidth]{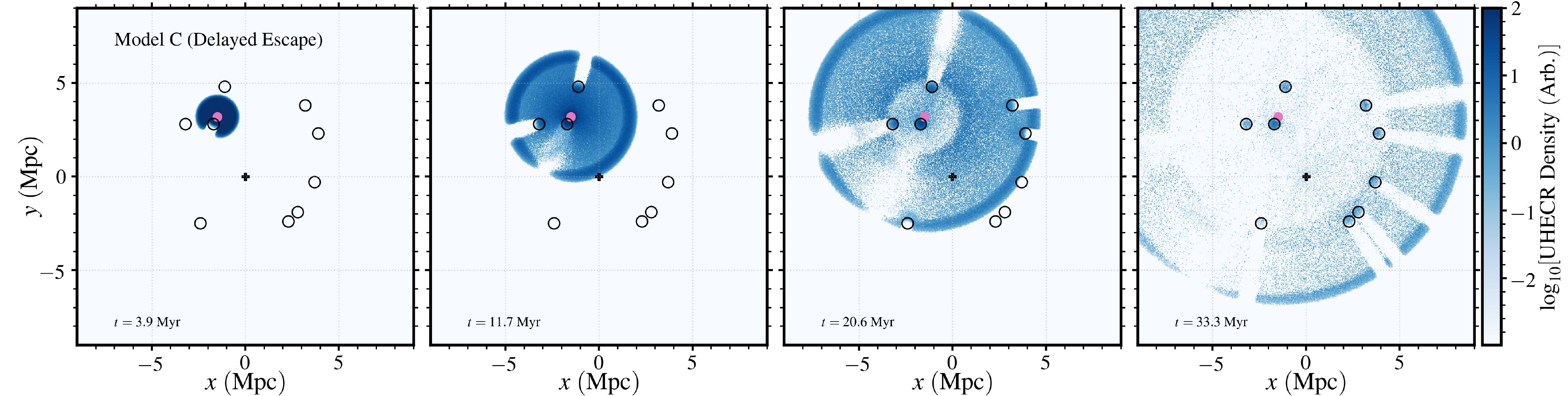}
\caption{Particle position maps from a slice of thickness $\Delta z = 0.6~{\rm Mpc}$ in the $z=0$ plane from Model~B and Model~C at four timesteps (3.9~\Myr, 11.7~\Myr, 20.6~\Myr, 33.3~\Myr), following their impulsive release from Cen~A. The position maps are presented as binned particle densities with bin sizes of $0.03~{\rm Mpc}$ and a density floor of $10^{-10}~{\rm bin}^{-1}$ in these arbitrary units. 
}
\label{fig:positionsbc}
\end{figure*}

In Fig.~\ref{fig:positionsbc}, snapshots of the logarithm (base 10) of the binned density is shown for the Model~B and Model~C simulations (bin size 0.03 Mpc), for four different times: 4~\Myr, 12~\Myr, 21~\Myr, and 33~\Myr . Also shown in this figure are the positions of the CoG objects (empty circles), Cen~A (pink filled circle), and the MW location (black vertical cross).

\end{document}